\documentstyle[11pt,aaspp4]{article}

\def\ga{\mathrel{\hbox{\rlap{\hbox{\lower4pt\hbox{$\sim$}}}\hbox{$>$}}}}
\def\la{\mathrel{\hbox{\rlap{\hbox{\lower4pt\hbox{$\sim$}}}\hbox{$<$}}}}
\def\kmsec{\hbox{${\rm km\ s^{-1}}$}}
\def\arcmin{\hbox{$^\prime$}}
\def\arcsec{\hbox{$^{\prime\prime}$}}

\def\leaderfill{\leaders\hbox to 1em{\hss.\hss}\hfill}

\lefthead{Ringwald and Naylor}
\righthead{Wind of the Cataclysmic Variable BZ Cam}

\begin{document}

\title{High-Speed Optical Spectroscopy of a Cataclysmic Variable Wind: BZ
Camelopardalis}

\author{F. A. Ringwald}
\affil{Department of Astronomy \& Astrophysics, The Pennsylvania State
University, \\ 525 Davey Laboratory, University Park, PA 16802-6305}
\authoremail{ringwald@astro.psu.edu}

\and

\author{T. Naylor}
\affil{Department of Physics, Keele University, Keele, Staffordshire,
ST5~5BG, U.\,K.}
\authoremail{timn@astro.keele.ac.uk}

\begin{abstract}

BZ Cam is the first cataclysmic variable star with an accretion disk wind
evident in its {\sl optical\/} spectrum.  The wind was found by
Thorstensen, who discovered intermittent P Cygni profiles occurring
simultaneously in He~I $\lambda$\,5876 \AA\ and H$\alpha.$ We have since
obtained spectra with 0.4-\AA\ pixel$^{-1}$ dispersion and 60-s time
resolution. 

We find a wind much faster and more rapidly variable than the radiatively
accelerated winds of OB stars, Wolf-Rayet stars, or luminous blue
variables.  Instead of showing blob ejection, the whole wind of BZ Cam
appears to turn on and off.  We use this to measure the acceleration law
of a CV wind for the first time.  The velocity increases linearly with
time, attaining blue edge velocities near $-3000$ \kmsec, and absorption
velocities near $- 1700$ \kmsec, in 6 to 8 min after starting near rest. 
We also find a subsequent linear deceleration to nearly rest in 30 to 40
min, perhaps an effect of dilution as the wind expands.  No periodicity
from rotational outflow is obvious. 

This wind is erratic and incessantly variable, and perhaps bipolar and
face-on, but not highly collimated.  The P Cygni absorption events trace
out sawtooth waves, occurring within 30 to 40 white dwarf radii from the
disk.  This is the approximate size of the disk, as well as the disk/wind
transition region recently postulated by Knigge and Drew.  We estimate a
distance of $830 \pm 160$~pc, and an orbital inclination $i$ such that
$12^{\circ} \la i < 40^{\circ}.$

\end{abstract}


\vfill\break

\section{Introduction}

\vbox{
\smallskip\noindent
``It is good to see you are keeping BZ.'' 

-- Prof.~J. R. Thorstensen}\medskip

Mass loss from cataclysmic variable binary stars (CVs) is nothing new.
Classical novae are well known to eject shells during their thermonuclear
eruptions.  CV winds were discovered with space ultraviolet spectra
(Krautter et al.~1981;  C\'ordova \& Mason 1982).  These winds' velocities
span 3000 -- 5000~\kmsec, the range of white dwarf escape velocities. 

BZ Cam {\it is\/} an unusual CV, however.  Long called just 0623+71, it
lies in a nebula, discovered by Ellis, Grayson, \& Bond (1984) and
designated EGB~4.  The nebula itself is unusual, having a bow-shock
morphology (Krautter et al.~1987; Hollis et al.~1992).  Also unusual about
BZ Cam is that its {\it optical\/} spectrum reveals a wind.  This was
serendipitously discovered by Thorstensen, Thomas, \& Patterson~(1993),
and published by Patterson et al.~(1996), or Paper I.  They found P Cygni
profiles, or blue-shifted absorption indicating outflows, occurring
intermittently in both the He~I $\lambda$\,5876 \AA\ and H$\alpha$
emission lines. 

BZ Cam does not show photometric behavior usually considered to be
outbursts (Garnavich \& Szkody 1988).  Such CVs are often called
nova-likes, because their spectra resemble those of novae between
eruptions---sometimes.  Nova-likes also resemble dwarf novae during their
accretion-powered outbursts, both having winds, for example.  CV winds are
thought to be bipolar, since the P Cygni profiles in their ultraviolet
resonance lines are not seen in eclipsing CVs (C\'ordova \& Mason 1985). 
BZ Cam does not show blue-shifted absorption clearly detached from the
main line profile, as in the symbiotic binary MWC 560 and attributed to
highly collimated jets (Tomov et al.~1992), but jets are rare in CVs,
anyway (Livio 1997; Shahbaz et al.~1997). 

Nearly all accretion disks, from those found in star formation to those of
AGNs, have outflows: see the review on CV winds by Drew (1997), as well as
many other papers on disk outflows in the same volume.  CVs are close
binary stars in which an approximately main-sequence K or M star fills its
Roche lobe and spills onto a white dwarf.  The orbital motion makes the
gas stream settle into a ring around the white dwarf.  Ill-understood
viscous dissipation spreads the ring into a disk, and gas accretes onto
the white dwarf.  CVs' luminosities are usually dominated by the light of
their accretion disks.  Sometimes, the white dwarf's magnetic field can
partially or completely disrupt the gas flow through the disk: such
systems are called magnetic CVs.  For reviews on magnetic CVs, see Cropper
(1990), Patterson (1994), and Warner (1995a), as well as many other
articles in this last volume. For recent reviews on CVs in general, see
Livio (1994) and Warner (1995b). 

In this paper, we present spectra of BZ Cam with five times the spectral
resolution and 12 times the time resolution of those of Paper I. These are
among the first high-speed spectra of the wind in any CV.  They show the
acceleration and apparent deceleration of a CV's wind for the first time.

\section{Observations}

We used the 4.2-m William Herschel Telescope and its ISIS double
spectrograph (Unger et al.~1988) on 1995 February 6 -- 7 and 7-- 8 UT, or
Nights 1 and 2.  The blue arm of ISIS was centered on He~I
$\lambda\,$5876~\AA, and the red arm on H$\alpha.$ We will refer to the
resulting spectra as the blue and the red spectra, respectively (even
though we know $\lambda$\,5876~\AA\ is yellow).  The R1200B and R1200R
gratings were used with TEK CCDs, with the dichroic centered on 6100 \AA. 
All spectra had 0.4 \AA\ pixel$^{-1}$ dispersion. 

The weather was photometric on Night 1, with seeing of 0.6 -- 0.8\arcsec\
throughout the run.  Some high clouds early on Night 2 were probably
avoided, as shown by the First Quarter Moon.  The slit was 1.0\arcsec\
wide on the sky.  It was set to a position angle of $173^{\circ},$ to
cover BZ Cam and a second star, 96\arcsec\ northwest, as a photometric
comparison. 

All exposures had 20-s integrations and 10 s of dead time, to read out the
CCDs. This yielded over 1300 spectra, total.  To improve the
signal-to-noise ratio (S/N), we co-added the spectra in pairs, so all
shown have 60-s time resolution.  Typical S/N at the continuum was 36
after co-adding, for all spectra. 

A problem we discovered with the data acquisition system haunted us
throughout the run.  Every time 400 spectra were taken, a bug deep in the
system software would awaken and stall it, demanding a reset of nearly all
the equipment.  This caused the larger gaps in the trailed spectrum plots
shown in Figures 1 and 3.  The smaller gaps were introduced by taking
spectra of Cu-Ne comparison lamps for wavelength calibration, as well as
of standard stars for flux calibration.  Lamps and standards were taken at
the beginning and end of each night, as well, and at the ends of the large
gaps.  The lamp spectra show the velocity scale was stable to within 5
\kmsec.  The Na~D lines, being interstellar (see Section 3.5), were
stationary to within 9.2 km~s$^{-1}.$ This reassures us that the
spectrograph and wavelength scale were stable.  Observing HD~32963 ($v_r =
-63.1 \pm 0.4$~\kmsec: Foster et al.~1997) allowed calibrating the
absolute velocity scale, to within $10.0 \pm 3.3$ \kmsec.  The Na~D lines
gave an average absolute heliocentric radial velocity, for all blue
spectra, of $+4.2 \pm 16.0$~\kmsec.  All velocities and times reported are
heliocentric, all times being for start of exposure.

\section{The spectra}

\subsection{Graphic representation}

In Figures 1 and 3, the time-resolved spectra have been stacked next to
each other, with the flux intensity depicted by the greyscale.  Emission
lines appear white, and absorption lines appear black.  All spectra were
rectified, to improve contrast in the plots.  The rectification fitted a
first-order cubic spline to each spectrum, and subsequently divided the
spectrum by it.  We chose this relatively simple function to avoid
removing genuine continuum features---or installing spurious ones, which
we carefully checked for by inspecting all spectra, before and after
rectification.  This was prompted by finding the unexpected line profile
phenomena described below, for example the emission to the blue of
H$\alpha$ (denoted as em2). 

The P Cygni events appear in Figures 1 and 3 as dark smudges to the left
of the emission line.  Diffuse white emission to the red of the main
H$\alpha$ emission line are features we will call red wings; they were
also present in the spectra of Paper I. The two black vertical stripes are
the Na~D lines.  Vertical striping in the grey continuum of the H$\alpha$
image is from flat-field division that should have been smoother. 

Figures 2 and 4 are schematic maps, included to help describe Figures 1
and 3.  Features are labeled in a system we have devised. As examples,
n1ba4 denotes Night 1 blue absorption 4, or the fourth P Cygni event on
Night 1 in the blue spectrum; n2rw3 denotes Night 2 red wing 3, or the
third thickening of the H$\alpha$ red wing on Night 2.

\subsection{The line components}

Velocities and equivalent widths of various line profile features are
plotted in Figures 5 and 6.  Figure 7 is a schematic key to explain our
notation for these features.  Many are similar to those of Prinja \& Rosen
(1995), but as we examined the spectra, we noticed more features.  Not all
spectra showed all the features shown in Figure 7, however.  Some do not
have red wings; some do not have the blue emission component (em2); many
don't even show P Cygni absorption. 

The dotted horizontal line in Figure 7 represents the normalized
continuum.  We define $V_{edge,blue}$ as the maximum blueward velocity of
any emission component, where it met the continuum.  Clearly, measurements
of $V_{edge,blue}$ depend on S/N.  $V_{em,cen}$ is the velocity of the
centroid of the main emission line component.  $V_{em2,cen}$ is the
velocity of the centroid of the second emission line component, to the
blue of the main component. $EW_{em}$ and $EW_{em2}$ are these components'
respective equivalent widths, including only area above the reference
continuum.  Figure 7 shows the limits they extend from and to, in all
cases where they met the continuum: at $V_{em2,red}$ to blue, and
$V_{em,blue}$ to red.  In no case did $EW_{em2} > EW_{em}.$

$V_{abs}$ is the velocity of the centroid of the absorption component; of
all these measured velocities, it is the one most clearly traced in
Figures 1 and 3.  $EW_{abs}$ is the equivalent width of this absorption
component, the area below the reference continuum.  $V_{redwing,blue}$
marks the local minimum in the emission line, in cases of a pronounced red
emission component. $V_{redwing,cen}$ is the centroid of the local
maximum, to the red of this local minimum.  $V_{em,red}$ represents the
maximum redward velocity of any emission component. 

Errors were estimated by count statistics (Horne, Wade, \& Szkody 1986). 
This ignores systematics, but a rigorous error analysis is impossible with
so many unknowns, e.g., how, where, and at what velocities the lines form.

\subsection{Spectrum variability}

At least eight definite P Cygni absorption events appeared, in a total of
6.6 hours of observations spread over 7.5 hours (see Figures 1 -- 4 and
Table~1).  This agrees with the rate of Paper I, with events recurring on
the average of about every two hours, and lasting for an average of about
one hour: about half the time, an event was occurring. 

CV winds have long been thought to resemble those of OB stars, because of
their similar temperatures and line profiles (Krautter et al.~1981; Mauche
\& Raymond 1987; Drew 1987, 1997). We were therefore surprised to find
that, except for an initial acceleration, the absorption troughs of the P
Cygni profiles of He~I $\lambda$\,5876 \AA\ moved redward over time, not
to the blue as expected for an outflow.  The maximum velocity of these
absorption troughs was about $-1700$~\kmsec\ from the rest frame.  The
events appeared from within 6 to 8 to as many as 20 min, and lasted for 30
to 40 min.  In both acceleration and deceleration, the velocities were
linear functions of time. 

As in Paper I, the P Cygni events showed no correlation with orbital
phase.  This is more obvious with He~I $\lambda$\,5876~\AA, since the
orbital velocity variation shows clearly in Figures 1 and 3, the
observations on both nights lasting just over one orbital period (3.679
hours, Paper I). The intervals between events were uneven: on Night 1,
events occurred near maximum blueshift and maximum redshift; on Night 2,
events occurred near descending node and maximum redshift.  Still, the
data have gaps, were taken on only two nights, and have short time
baselines modulo one day.  A long, multi-longitude, time-resolved
spectroscopic study may yet show periodic behavior. 

H$\alpha$ also showed P Cygni events, occurring simultaneously with those
of He~I $\lambda$\,5876~\AA.  The red wing of H$\alpha$ extended by up to
$+2400$~\kmsec\ on both nights.  This red-shifted emission may indicate
rotating outflow, being a complement to blue-shifted absorption (Drew
1987; Mauche \& Raymond 1987).  However, it was uncorrelated with the P
Cygni events: n1ra2 and n1rw1 occurred simultaneously, but n1ra1, n2ra1,
and n2ra3 are not associated with an extended red wing.  Red wing events
can appear and disappear quite suddenly, within minutes. 

There may be structure in both lines (see Figure 8), in spectra 2 and 12
of Night 1 in He~I, and in spectrum 2 of Night 2 in He~I.  Perhaps this is
blob ejection, but this is quite unlike the discrete absorption components
in hot star winds, being small perturbations for example, in the O7.5
giant $\xi$ Persei (Prinja, Howarth, \& Henrichs 1987), or in Wolf-Rayet
stars (Moffat et al.~1988), or in the central stars of planetary nebulae
(M\'endez et al.~1990; Balick et al.~1996). In BZ Cam, the whole wind
turns on and off.  This is somewhat reminiscent of luminous blue
variables, but on a much faster timescale, and at much higher speeds,
since LBV winds vary over hundreds of \kmsec\ over hundreds of days (Wolf
1992). 

Included in Figures 5 and 6 for H$\alpha$ are velocities measured by the
method of Schneider \& Young (1980), of convolving the line profiles with
the derivative of a Gaussian and taking the centroids as the velocities. 
A Gaussian of 275 km~s$^{-1}$ FWHM was used, to allow comparison with
Paper I.  These velocities were then fit by least squares to a sinusoid of
period held fixed at 0.1533 days.  The fit is terrible (see Table~2), but
then Paper I had similar problems, due to line profile variations. A more
appropriate ephemeris would use velocities less disturbed by this: we
therefore adopted $V_{em,cen}$ for H$\alpha,$ which best fits a sinusoid,
and based the ephemerides in this paper on it.  This is of course just a
relative ephemeris, measured from emission lines, and so not necessarily
tracing the motion of either star (e.g., Thorstensen et al.~1991).

\subsection{He~I $\lambda$\,6678~\AA, inclination, and opening angle}

He~I $\lambda$\,6678~\AA\ traces a sinusoid as it moves about the orbit. 
Throughout Figure 8, it may show doubling in its line profile, but the
line is faint and noisy.  Nevertheless, whenever the line does appear
doubled, the two peaks have a separation of $220 \pm 30$~\kmsec, or about
12 pixels. 

Comparing this to calculated line profiles (Horne \& Marsh 1986), we
estimate the orbital inclination $i \approx 12 \pm 3^{\circ}.$ This should
be considered a lower limit, since BZ Cam has a strong wind, which could
easily blur the line profile.  One may alternatively assume the peaks come
from the disk's edge (Shahbaz et al.~1997), so the projected velocity is
$v_{disk} \sin i = 110$~\kmsec. The Keplerian velocity of the disk edge is
$v_{disk} = 1081 [M_1/P_{orb}(h)]^{0.5},$ where $M_1$ is the white dwarf
mass in $M_{\odot},$ and $P_{orb}(h)$ is the orbital period in hours. 
Throughout the range of plausible white dwarf masses, from 0.3 to 1.4
$M_{\odot},$ the predicted $i$ varies only from $> 12.3^{\circ}$ to $>
7.3^{\circ},$ respectively.  A weak upper limit may be estimated by
assuming an average wind speed of 4000~\kmsec. With the observed maximum
$V_{em,blue} \sim 3000$~\kmsec, and $\cos i \sim
({V_{em,blue}}/V_{wind}),$ we estimate $i < 40^{\circ}.$ Surely $i <
70^{\circ},$ with no eclipses. 

The opening angle $\theta$ of the wind may be estimated by $\sin \theta =
\Delta V/(V \tan i),$ where $\Delta V$ is the range in radial velocity of
the emission components with respect to their mean velocities. From the
widths of the red wing and em2 features, we estimate $\Delta V/V \sim
0.5.$ This implies $\theta > 37^{\circ},$ for $i < 40^{\circ}.$ Surely
$\theta > 10^{\circ},$ since $i < 70^{\circ}.$ If $i < 26^{\circ},$ there
is no restriction on the wind's opening angle.

\subsection{Na~D, distance, and $M_V$}

A CV with BZ Cam's orbital period should have an approximately dM4
secondary star (Patterson 1984).  Such a star should have $M_V = 12.7$
(Bessell 1991), over 9 magnitudes fainter than the system (see below).
Even with an equivalent width of 20 \AA\ for Na~D, typical for a dM4 star
(Montes \& Martin 1997), the secondary's contribution to Na~D should be
negligible.  There may be some contribution from cooler parts of the disk,
but the Na~D lines were stationary (see Section 2), and so largely
interstellar.  We expect absorption from the nebula to be negligible,
since although Hollis et al.~(1992) found the nebula to be 35 times denser
than the surrounding ISM, the pathlength through the nebula is $< 0.06$\%
the pathlength to Earth, the nebula subtending only 2\arcmin\ (Krautter,
Klaas, \& Radons 1987). 

The average equivalent widths of both Na~D lines was 0.42 $\pm$ 0.08 \AA. 
With the relation of Allen (1973), we deduce a distance for BZ Cam of $830
\pm 160$~pc, versus 500~pc assumed by Krautter et al.~(1987) and by Hollis
et al.~(1992). 

BZ Cam varies over $12.5 < m_{pg} < 14.0,$ with an average $m_{pg} = 13.0$
(Garnavich \& Szkody 1988). From through-the-slit magnitudes, we estimate
that during our observations, $V \approx 12.7 \pm 0.2,$ with variability
less than the errors.  This brightness was similar to that in Paper I of
$V = 12.5 \pm 0.2$.  We assume $m_{pg} \approx B$ and $B - V = 0.0,$
typical for nova-likes (Bruch \& Engel 1995), and $E(B - V) = 0.05 \pm
0.04$ (Verbunt 1987) with $A_V = 3.20 E(B - V)$ (Seaton 1979). 
Paczy{\'n}ski \& Schwarzenberg-Czerny (1980) give an inclination
correction for an optically thick disk $\Delta M_V (i) = -2.5 \log [(1 +
{3 \over 2} \cos i) \cos i]$ = -0.96 for $i = 12^{\circ},$ and negative
for $i < 56.7^{\circ}.$ For a distance of 830~pc, this implies an average
$M_V \ga 4.3,$ ranging over $3.8 < M_V < 5.3$---among the most luminous
CVs, typically near $M_V = 4.2$ (Warner 1987).

\section{Discussion}

\subsection{Is BZ Cam really a CV?}

Two aspects of BZ Cam are unusual, for a CV: its nebula and its
line-profile variability.  One should therefore consider carefully whether
BZ Cam really is a CV, and whether it has an accretion disk. 

BZ Cam is certainly a close binary star system, as shown by its
sinusoidally variable radial velocity. BZ Cam shows many hallmarks of a
CV, since it flickers, its orbital period is 3.679~h, and its spectrum
resembles that of many nova-like CVs, with relatively weak lines on a
strong continuum (Williams 1983; Paper I). 

Both disk and magnetic CVs show erratic flickering in their light curves,
thought to be from accretion, but still poorly understood.  Still, it may
be possible that BZ Cam is a detached close binary, and its observed
flickering is not that of a CV, but is instead the erratic,
``semi-irregular'' variability of planetary nebula nuclei with strong
winds (Bond \& Ciardullo 1989; Handler et al.~1996).  The observed
photometry of BZ Cam, however, varies on the order of tens of seconds or
faster (Pajdosz \& Zo{\l}a 1992; Paper I).  This seems too rapid for a
planetary nucleus, which varies over hours.  Photometry of planetary
nuclei could have shown faster variability from flickering, but did not:
it did show unrelated non-radial pulsations occurring over minutes
(Ciardullo \& Bond 1996). 

Lines in semi-irregular planetary nuclei do show radial velocity
variations.  These generally have amplitudes of $< 100$~\kmsec\ (M\'endez,
Herrero, \& Manchado 1990), though, much smaller than in BZ Cam. However,
the WC8 nucleus of NGC~40 has spectroscopic flickering in the center of
its C~III $\lambda$\,5696~\AA\ emission line that occurs over hours, with
amplitudes of $\sim 1000$ \kmsec---which interestingly, change in velocity
linearly with time (Balick et al.~1996).  Hollis et al.~(1992) argue that
EGB~4 is not a planetary nebula, because they find the
uncharacteristically low mass of $2.4 \times 10^{-3} M_{\odot}$, and
because it is entirely shock-ionized. 

BZ Cam may or may not have an accretion disk.  The direct evidence of a
disk is scanty: 0.8-\AA\ resolution spectra should have shown a disk's
distinctive double-peaked profiles in the emission lines (Honeycutt,
Kaitchuck, \& Schlegel 1987).  Other nova-likes with orbital periods
similar to BZ Cam's, the dreaded SW Sex stars (Thorstensen et al.~1991;
Hellier 1996), also show single-peaked emission, even when they have deep
eclipses.  Our spectra of BZ Cam show no clear cases of double peaks in
H$\alpha,$ but as mentioned in Section 3.4, the He~I $\lambda$\,6678~\AA\
line had double peaks in about half the spectra containing this line. 

The spectra of Williams (1983) were all taken at the same resolution, and
so invite comparison.  They show BZ Cam with a strong continuum, with
H$\alpha,$ the strongest emission line, having equivalent width of only
$-6.5$~\AA.  The H$\gamma$ and He~I lines were in weak absorption, with
equivalent widths ranging from 3 to 12~\AA.  The Balmer jump was in
absorption, and there was little high-excitation emission from He~II
$\lambda$\,4686~\AA.  In contrast, Williams's spectra of magnetic CVs show
H$\alpha$ equivalent widths ranging from 16 to 87 \AA, other Balmer lines
and Balmer jumps invariably in emission, and He~II $\lambda$\,4686~\AA\
emission lines among the strongest in the spectra.  Magnetic CVs also do
not show P Cygni profiles in their ultraviolet spectra.  (An exception may
be 1H~0551$-$819:  Mouchet et al.~1996.) Disk CVs often do, as does BZ Cam
(Woods, Drew, \& Verbunt 1990; Griffith, Fabian, \& Sion 1995).  Magnetic
CVs show red wings in H$\alpha$ (Crosa et al.~1981), but these are
attributed to the gas stream, since they vary with orbital phase, unlike
in BZ Cam.  BZ Cam also does not show the coherent optical pulsations of
magnetic CVs (Pajdosz \& Zo{\l}a 1992;  Paper I).  Polarimetry, another
indicator of magnetic CVs, appears not to have been published. 

To summarize, BZ Cam is definitely a close binary.  That it is anything
but a CV seems improbable. It is likely to have a disk, although the
evidence for this is not stunning.  Still, the evidence for the
alternative, that it is magnetic, is worse.  The low $K$-velocity (Table
2; Paper I) and narrow emission line widths suggest an almost face-on
accretion disk.  This may explain why we see P Cygni events, since nearly
all accretion disks seem to be attended by bipolar outflows (e.g., Livio
1997).  As shown in Section 3.5, its absolute magnitude appears to be
brighter than average for a CV, but not astoundingly so.  The simplest
solution that fits all properties is that BZ Cam is a nearly face-on CV
with a high-than-average mass transfer rate.

\subsection{Mass loss and the acceleration of a CV wind}

Why does BZ Cam do this?  Only now is evidence for mass loss in CVs
accumulating from their optical spectra, including AT Cnc, which showed a
P Cygni profile in H$\alpha$ (Smith et al.~1997), and V1315 Aql, which
Hellier (1996) argues to show a P Cyni profile convolved onto its lines,
although it is an eclipsing SW Sex star.  V1193 Ori also at least once
showed an H$\alpha$ red wing, not unlike that in BZ Cam (Ringwald,
Thorstensen, \& Hamwey 1994).  It would be of interest to check these CVs'
fields for faint nebulae. 

Still, if this were common behavior for CVs, might it not have been seen
long ago, in the hundreds of radial velocity studies done of CVs (e.g.,
Kraft 1964; Ritter \& Kolb 1997)?  It may partly be because BZ Cam may be
nearly face-on, but this behavior has not been noticed around other
bright, face-on, luminous CVs, such as RW Sex (Beuermann, Stasiewski, \&
Schwope 1992) or the old nova V603 Aql (Patterson et al.~1997).  We note
there is no published time-resolved spectroscopy of the bright VY Scl star
MV Lyr while in its high state. 

If there is a connection between high disk mass transfer rate and high
wind mass transfer rate, one might surmise that BZ Cam has an high disk
mass transfer rate, as corroborated by its high luminosity.  We therefore
hypothesize that BZ Cam shows P Cygni events and red wings because it has
a roughly bipolar outflowing wind, seen face-on, although a spherical
outflow cannot be excluded.  Ejected shells coming toward Earth give the
lines P Cygni profiles.  When they expand and become dilute, they turn
into emission lines, which disappear when they get faint.  Ejected shells
moving away from Earth manifest themselves as redshifted emission. 
Intermittent redshifted absorption features are not seen because they are
blocked by the accretion disk. 

This implies that BZ Cam has a disk.  The blocking cannot be from an
optically thick part of blue-shifted wind, since it is not always present
when the red wing is (shown by events n1rw2 and n2rw2: see Figures 1 --
4).  H$\alpha$ is a lower excitation line than He~I $\lambda$\,5876~\AA,
which would explain why its blue-shifted absorption turns into
blue-shifted emission as the wind expands, and why that of He~I
$\lambda$\,5876~\AA\ does not. 

That the apparent deceleration of the wind may be due to dilution as it
expands is supported by the simultaneous behavior of He~I
$\lambda$\,5876~\AA\ and H$\alpha$ in nearly all events.  The shapes of
all these events in the trailed spectrum plots are roughly the same in
both lines, but H$\alpha$ switches to emission when He~I stays in
absorption. 

Since the flow occurs on characteristic timescales and velocity scales,
one may therefore estimate the length over which the wind accelerates. 
Assuming constant acceleration from rest perpendicular to a face-on disk,
$r = k t^2 /2,$ where $k = 1700$~\kmsec\ in 6 -- 8 min, or $(3.5 \rm{-} 5)
\times 10^5$~cm~s$^{-2},$ and $r \sim (3 \rm{-} 4) \times 10^{10}$~cm. 
This is comparable to estimated disk radii in luminous CVs (Sulkanen,
Brasure, \& Patterson 1981; Harrop-Allin \& Warner 1996).  At $R_{accel} =
30 - 40~R_{WD},$ with a white dwarf radius $R_{WD} = 0.014 R_{\odot},$ it
is also comparable to the $40 R_{WD}$ size of the disk/wind transition
region postulated by Knigge et al.~(1997) and Knigge \& Drew (1997). 

Figures 1 and 3 show that the P Cygni events trace out sawtooth waves in
time, suggesting the presence of shocks.  Are we seeing shocks in a
transition region, then?  Instabilities in hot star winds were unexpected
and are still poorly understood.  Another source of the random variation
of the P Cygni profiles may possibly be magnetism, long known to power the
apparently random, rapid activity in stellar flares and prominences.

\acknowledgments 

F. A. R. thanks PPARC for travel funding. T. N. holds a PPARC Advanced
Fellowship. The William Herschel Telescope is operated on La Palma by the
Royal Greenwich Observatory at the Spanish Observatorio del Roque de los
Muchachos of the Instituto de Astrofisica de Canarias. This work has used
the Simbad database, operated at CDS, Strasbourg, France.  Some analysis
was done with the ARK software on the Keele STARLINK node.  Some was done
with IRAF, which is distributed by the National Optical Astronomy
Observatories, operated by the Association of Universities for Research in
Astronomy, Inc.~(AURA), under cooperative agreement with the National
Science Foundation.  Special thanks are due to William A. Kramer, for
donating another computer on which analysis was done.  We also thank the
David Dunham Foundation for a Theodore Dunham, Jr.~Grant, the American
Astronomical Society for a Small Research Grant, Mark Somers for
assistance on and after the observing run, Raman Prinja for helpful
discussions, and Richard Wade for thought-provoking comments.



\begin{figure}
 \epsscale{0.8}
 \plotone{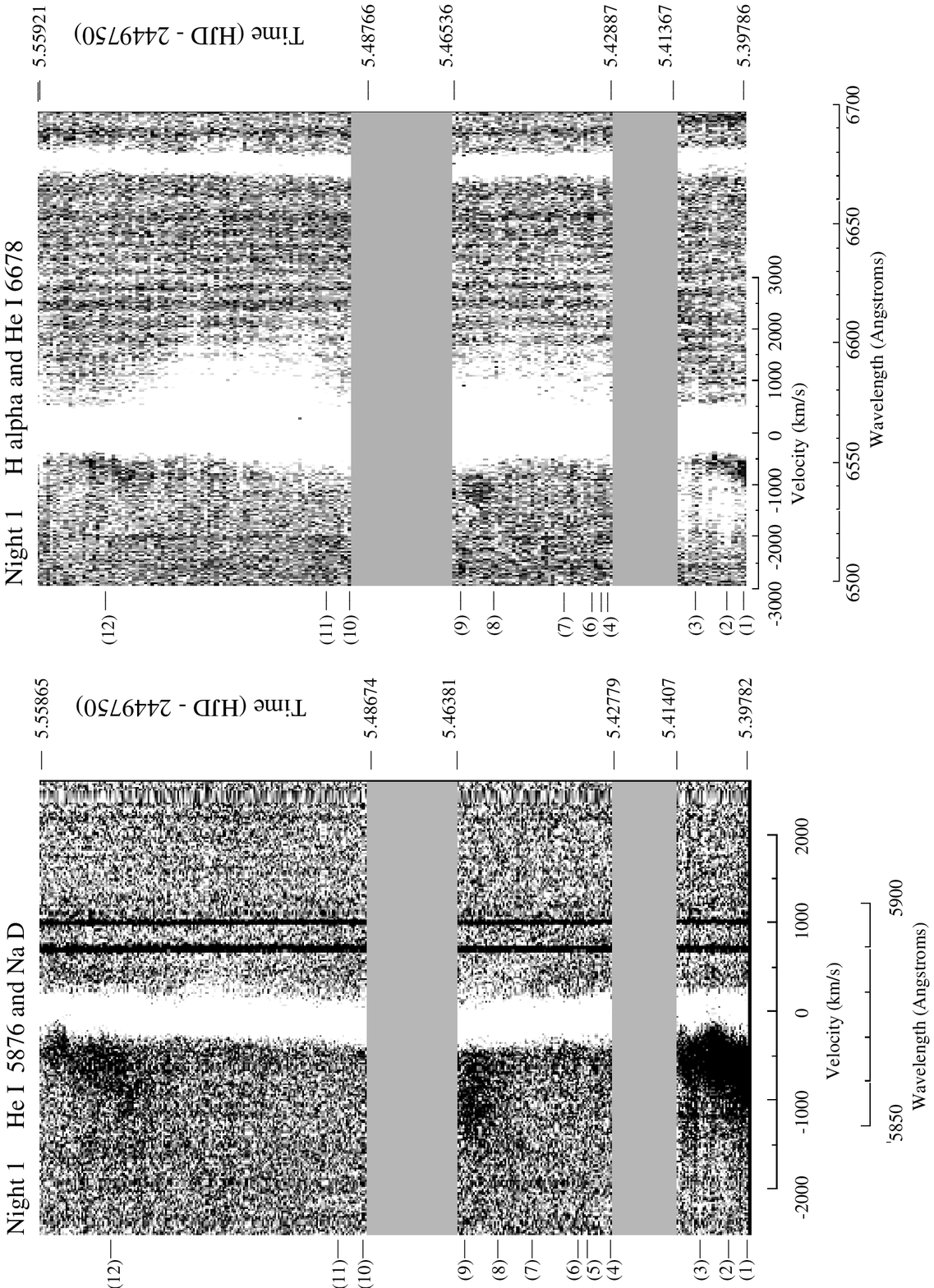}
\caption{
Greyscale rectified spectra of BZ Cam for 1995 February 6 -- 7 UT.  The
greyscale shows intensity:  emission lines are white, and absorption lines
are black.  Intermittent P Cygni events are black smudges to the left of
the main emission line; red wings are diffuse white emission.  Numbers in
parentheses at left indicate corresponding spectra in Figure
8.\label{fig1}}

\end{figure}

 \clearpage

\begin{figure}
 \epsscale{0.8}
 \plotone{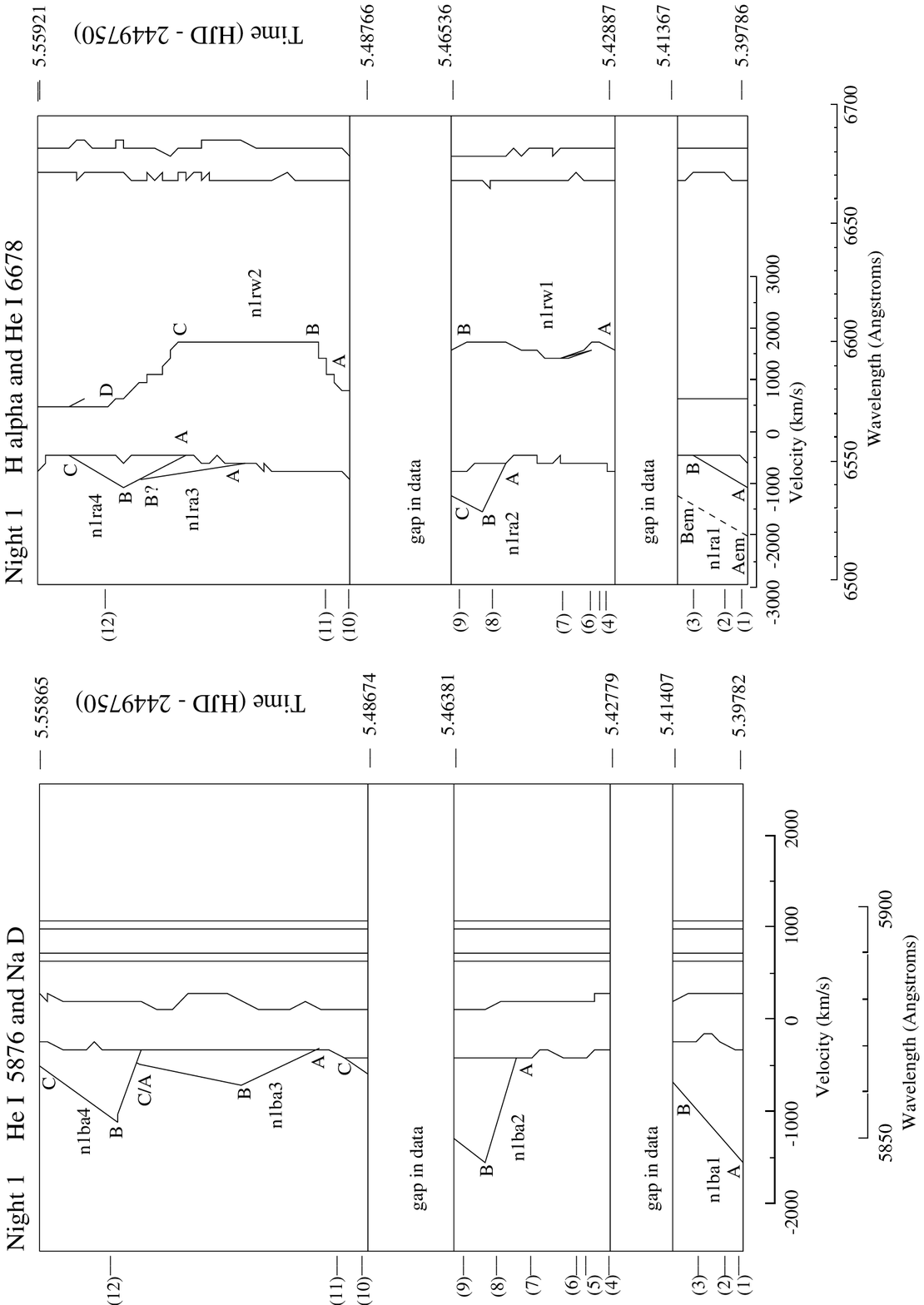}
\caption{
Schematic map for Figure 1, showing labeled features of interest. Table 1
lists corresponding velocities and times.  The series spans 1.05 
orbits.\label{fig2}}

\end{figure}

 \clearpage

\begin{figure}
 \epsscale{0.8}
 \plotone{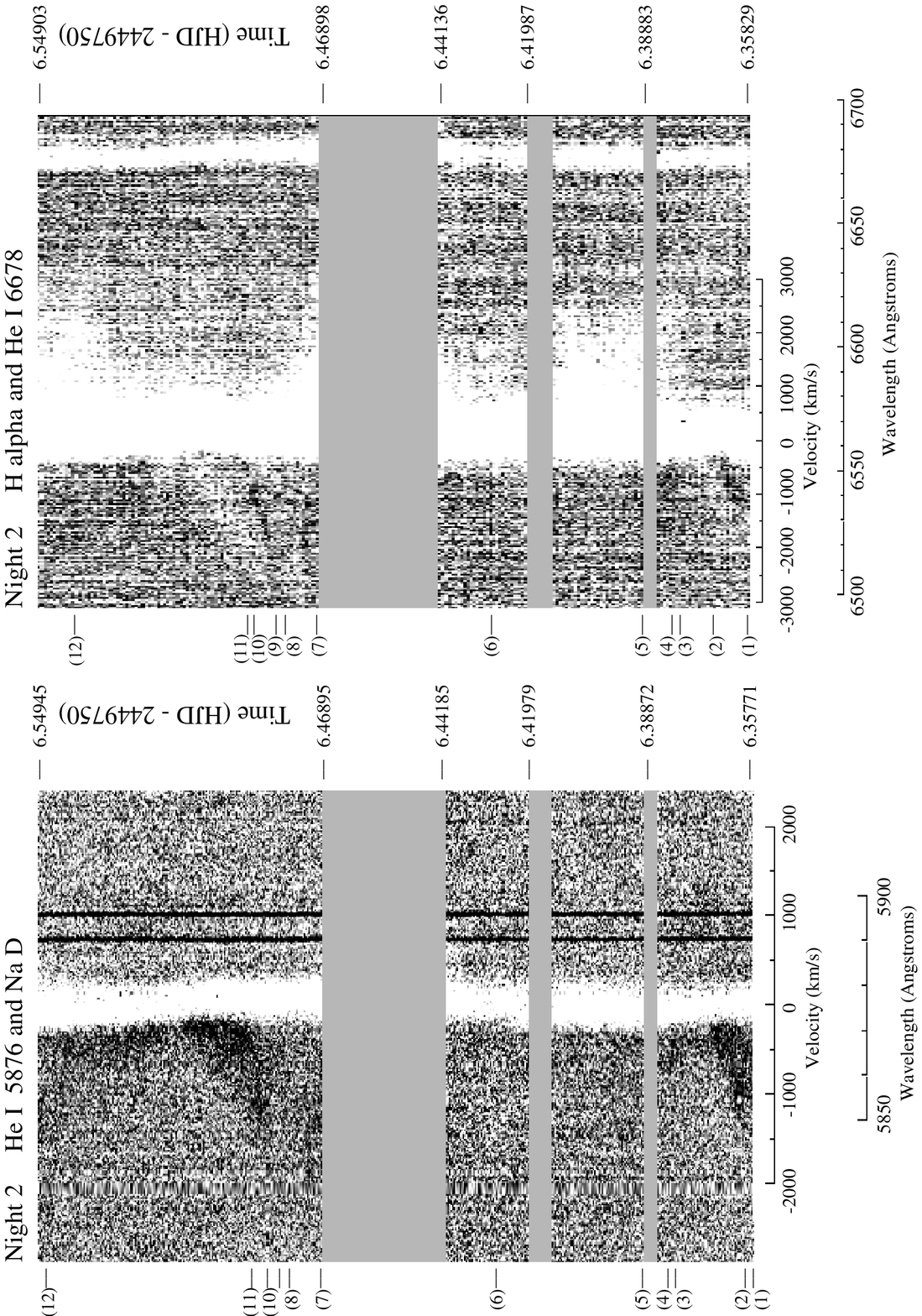}
\caption{
Greyscale rectified spectra of BZ Cam for 1995 February 7 -- 8 UT.  The
greyscale shows intensity: emission lines are white, and absorption lines
are black.  Intermittent P Cygni events are black smudges to the left of
the main emission line;  red wings are diffuse white emission.  Numbers in
parentheses at left indicate corresponding spectra in Figure
8.\label{fig3}}

\end{figure}

 \clearpage

\begin{figure}
 \epsscale{0.8}
 \plotone{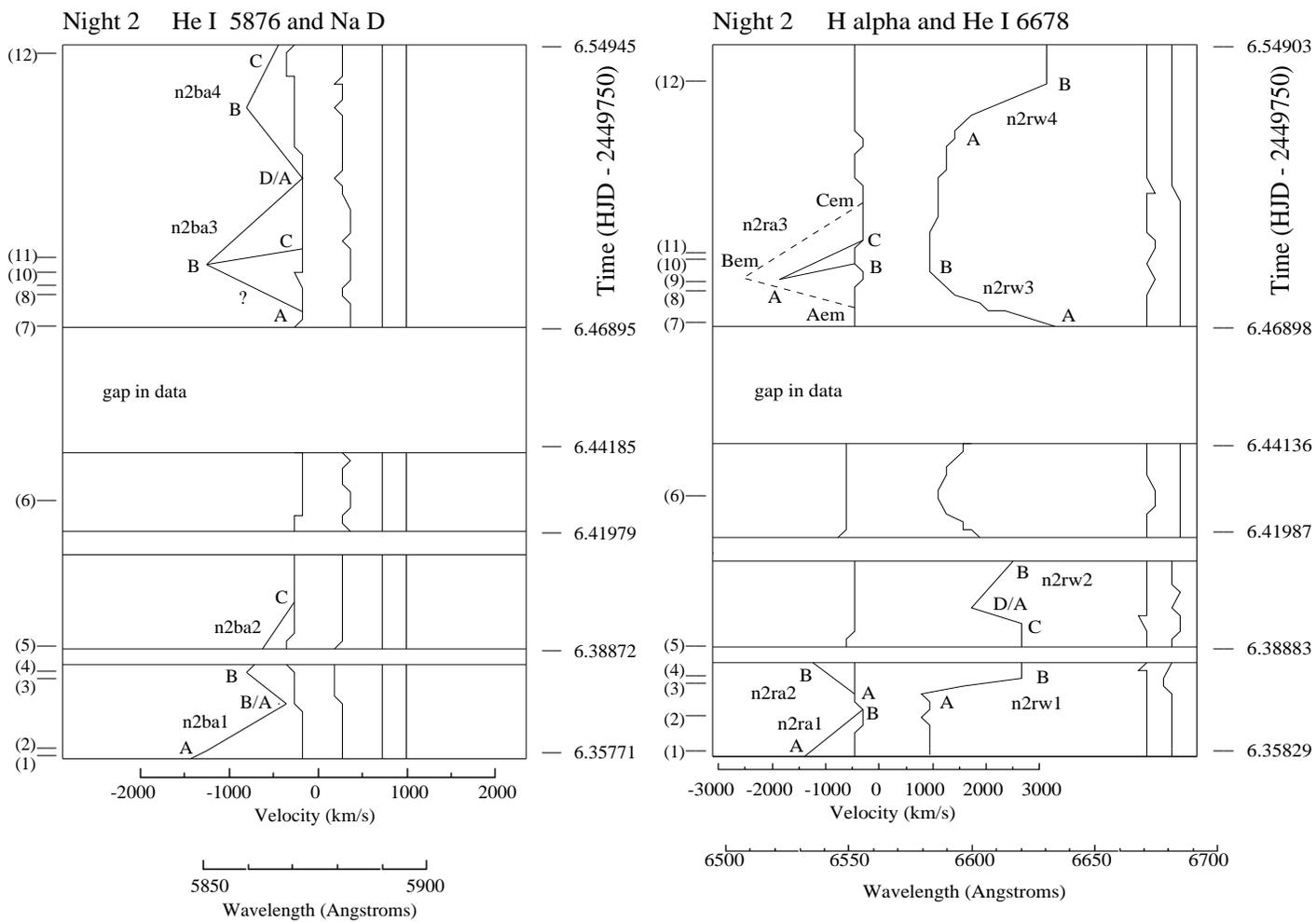}
\caption{
Schematic map for Figure 3, showing labeled features of interest. Table 1
lists corresponding velocities and times.  The series spans 1.25
orbits.\label{fig4}}

\end{figure}

 \clearpage

\begin{figure}
 \plotfiddle{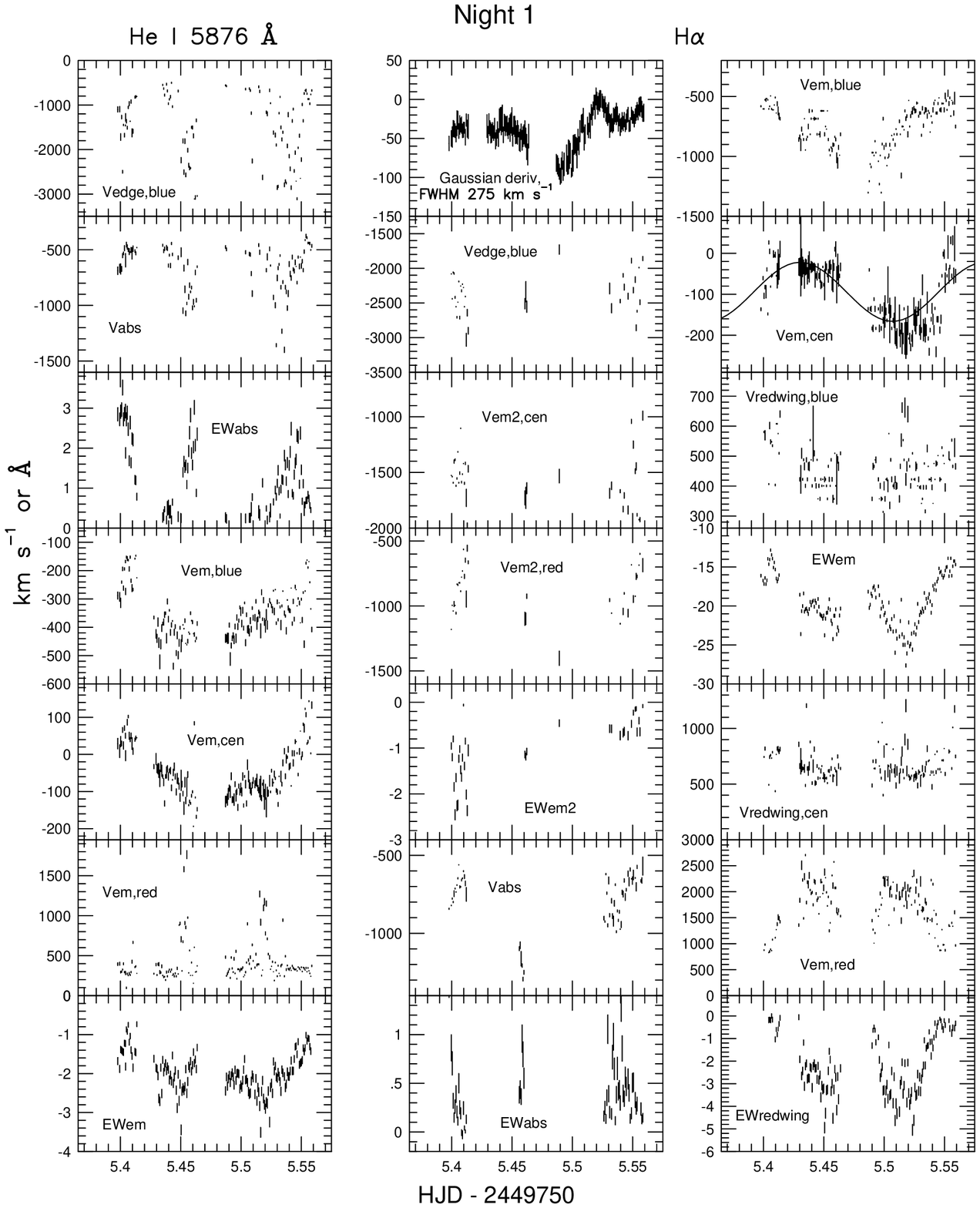}{7.75in}{0}{85.}{85.}{-260}{-45}
 \vskip-0.75in
\caption{
Measurements for Night 1 line profile features, shown in Figure 7 and
described in Section 3.2.\label{fig5}}

\end{figure}

 \clearpage

\begin{figure}
 \plotfiddle{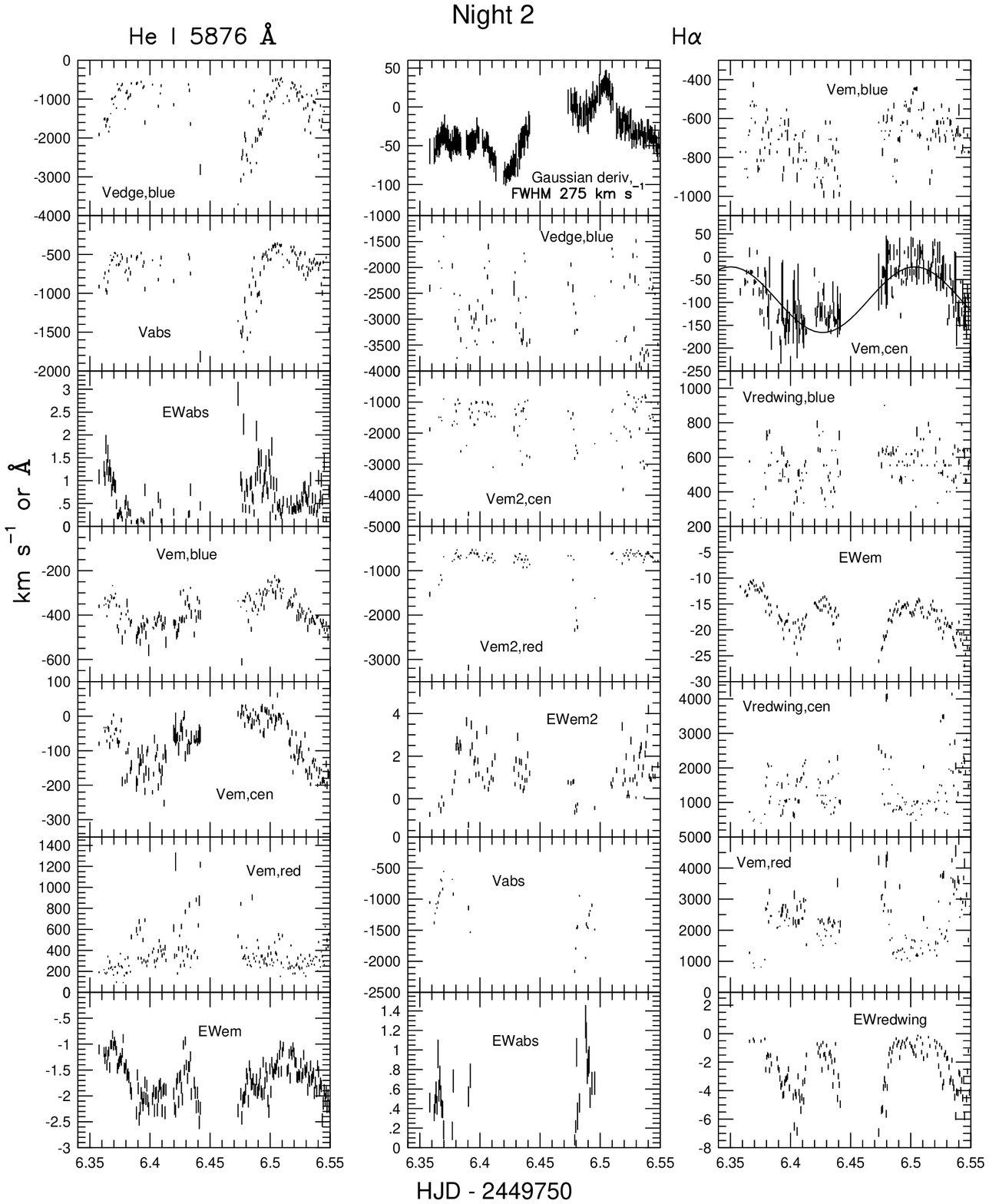}{7.75in}{0}{85.}{85.}{-260}{-45}
 \vskip-0.75in
\caption{
Measurements for Night 2 line profile features, shown in Figure 7 and
described in Section 3.2.\label{fig6}}

\end{figure}

 \clearpage

\begin{figure}
 \epsscale{0.6}
 \plotone{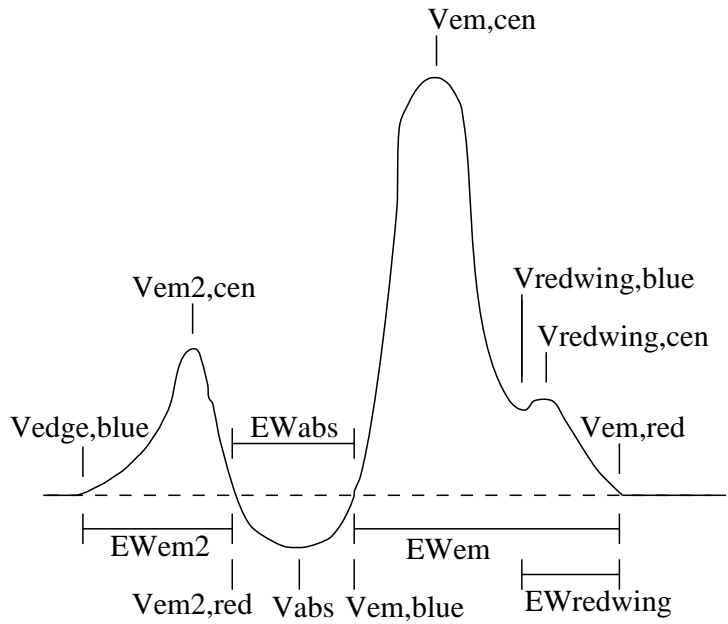}
\caption{
Schematic key to measured line profile features plotted in Figures 5 and
6.  This includes all phenomena noticed in the line profiles, although
almost no spectra showed all these phenomena together.\label{fig7}}

\end{figure}

 \clearpage

\begin{figure}
 \plotfiddle{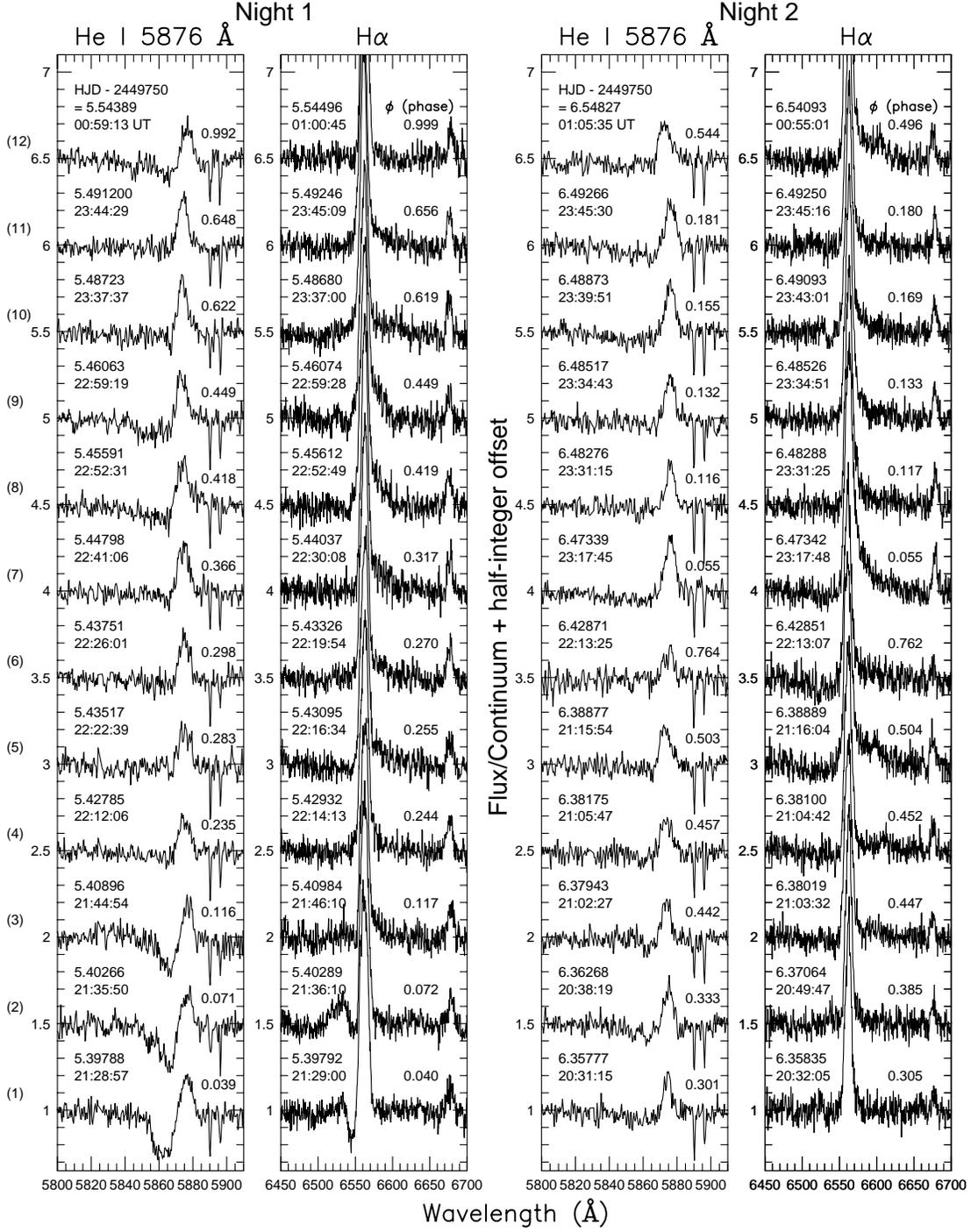}{7.75in}{0}{80.}{80.}{-245}{0}
 \vskip-0.75in
\caption{ 
Selected individual rectified spectra of BZ Cam.  Numbers in parentheses
at left indicate corresponding spectra in Figures 1 and 3.  Relative
orbital phase ${\phi}$ assumes $P_{orb} = 0.1533~d$ and $\phi = 0$ at $HJD
= 2449755.392,$ derived from $V_{em,cen}$ of H$\alpha$ (see Section 3.2
and Table 2).\label{fig8}}

\end{figure}

\clearpage

\parindent 0pt
\parskip 0pt
\def\tablerule{\noalign{\hrule\smallskip}}
$$\vbox{\halign{
# \hfil & # \hfil & # \hfil & \hfil # & $\;$ # &
# \hfil & # \hfil & # \hfil & \hfil # & $\;$ # &
# \hfil & # \hfil & # \hfil & \hfil # \cr
\multispan{14}{\hfil {\bf{Table 1. Chronology of Nights 1 \& 2 (Figures 1 
-- 4)}} \hfil} \cr \noalign{\smallskip} \cr \tablerule \cr \tablerule \cr
Event & \hfil $HJD{^a} -$ & \hfil ${\phi}^b$ & $v^c$ \hfil & &
Event & \hfil $HJD{^a} -$ & \hfil ${\phi}^b$ & $v^c$ \hfil & &
Event & \hfil $HJD{^a} -$ & \hfil ${\phi}^b$ & $v^c$ \hfil \cr
      & \hfil $2449750$   &            &       & &
      & \hfil $2449750$   &            &       & & 
      & \hfil $2449750$   &            & \cr
\noalign{\smallskip}
\noalign{\smallskip} \tablerule \cr
Night 1 & & \hidewidth{He~I $\lambda$ 5876 \AA}\hidewidth & & & & 
\hfil {H$\alpha$} & & & & & {H$\alpha$ red wing}\hidewidth & & \cr
 & & & & & n1ra1Aem & 5.39786 & 0.039 & $-2133$ & & & & & \cr
 & & & & & n1ra1Bem & 5.41367 & 0.142 & $-1941$ 
& & & & & \cr
n1ba1A & 5.39782 & 0.039 & $-1104$ 
& & n1ra1A & 5.39786 & 0.039 & $-1182$
& & & & & \cr
n1ba1B & 5.41407 & 0.147 & $-697$ 
& & n1ra1B & 5.41367 & 0.142 & $-798$ &
& & & & \cr
n1ba2A & 5.44557 & 0.350 & $-372$ 
& & n1ra2A & 5.44610 & 0.354 & $-506$ &
& n1rw1A & 5.42887 & 0.242 & $+683$\cr
n1ba2B & 5.45756 & 0.429 & $-1612$ 
& & n1ra2B & 5.45948 & 0.441 & $-1493$ & 
&        &         &       &       \cr 
n1ba2C & 5.48816 & 0.628 & $-403$ 
& & n1ra2C & 5.46536 & 0.480 & $-1338$ &
& n1rw1B & 5.46536 & 0.480 & $+875$\cr 
n1ba3A & 5.49229 & 0.655 & $-393$ 
& & & & &
& & n1rw2A & 5.48766 & 0.625 & $+518$\cr
n1ba3B & 5.50511 & 0.739 & $-829$ 
& & n1ra3A & 5.50793 & 0.757 & $-543$
& & n1rw2B & 5.49861 & 0.631 & $+1725$\cr
n1ba3C & 5.53612 & 0.941 & $-504$ 
& & n1ra3B? & 5.53611 & 0.925 & $-1018$ & 
&        &         &       &       \cr 
n1ba4A & 5.52289 & 0.855 & $-321$ 
& & n1ra4A & 5.53205 & 0.915 & $-533$
& & n1rw2C & 5.52496 & 0.868 & $+1661$\cr
n1ba4B & 5.53922 & 0.961 & $-1195$ 
& & n1ra4B & 5.53894 & 0.960 & $-1109$
& & n1rw2D & 5.53955 & 0.963 & $+482$\cr
n1ba4C & 5.55865 & 0.088 & $-514$ & & n1ra4C & 5.55536 & 0.067 & $-479$ & 
&        &         &       &       \cr 
\noalign{\smallskip}
\noalign{\smallskip} \tablerule \cr
Night 2 & & \hidewidth{He~I $\lambda$ 5876 \AA}\hidewidth & & & & 
\hfil {H$\alpha$} & & & & & {H$\alpha$ red wing}\hidewidth & & \cr
\noalign{\smallskip}
n2ba1A & 6.35771 & 0.300 & $-1855$ 
& & n2ra1A & 6.35829 & 0.304 & $-1512$ 
& & n2rw1A & 6.37654 & 0.423 &  $+536$\cr
n2ba1B & 6.36975 & 0.379 &  $-321$ 
& & n2ra1B & 6.37052 & 0.384 &  $-516$ 
& & n2rw1B & 6.37874 & 0.438 & $+2218$\cr
n2ba2A & 6.36975 & 0.379 &  $-321$ 
& & n2ra2A & 6.37373 & 0.405 &  $-543$ 
& & n2rw1C & 6.39178 & 0.523 & $+2236$\cr
n2ba2B & 6.38009 & 0.446 &  $-951$ 
& & n2ra2B & 6.38275 & 0.464 & $-1119$ 
& & n2rw1D & 6.39699 & 0.557 &  $+609$\cr
n2ba2C & 6.39826 & 0.565 &  $-382$ 
& & & & & 
& & n2rw2A & 6.39699 & 0.557 &  $+609$\cr 
n2ba3A & 6.47533 & 0.068 & $-301$ 
& & n2ra3Aem & 6.47219 & 0.047 & $-497$
& & n2rw2B & 6.41003 & 0.642 & $+2328$\cr
n2ba3B & 6.48779 & 0.149 & $-1825$ 
& & n2ra3A & 6.47921 & 0.093 & $-2664$
& & n2rw3A & 6.46898 & 0.026 & $+3096$\cr
n2ba3C & 6.49138 & 0.172 & $-280$ 
& & n2ra3Bem & 6.47801 & 0.085 & $-3048$
& & n2rw3B & 6.47912 & 0.092 &  $+609$\cr
n2ba3D & 6.51187 & 0.306 & $-291$ 
& & n2ra3B & 6.48643 & 0.140 & $-525$ 
& & n2rw4A & 6.51992 & 0.359 &  $+637$\cr
n2ba4A & 6.51187 & 0.306 & $-291$ 
& & n2ra3C & 6.49846 & 0.219 & $-424$ 
& & n2rw4B & 6.52834 & 0.414 & $+2630$\cr
n2ba4B & 6.52665 & 0.402 & $-951$ 
& & n2ra3Cem & 6.50989 & 0.293 & $-452$ 
& & & & & \cr
n2ba4C & 6.54945 & 0.551 & $-707$ & \cr
\noalign{\medskip} \tablerule \cr \noalign{\hrule} \cr
\noalign{\vskip 1pt} 
\noalign{$\strut^{\rm a}\ $Heliocentric Julian Date of the start of
integration} 
\noalign{$\strut^{\rm b}\ $Relative orbital phase, with $P_{orb} =
0.1533$~d and $\phi = 0$ at $HJD = 2449755.392$ (see Section 3.2)}
\noalign{$\strut^{\rm c}\ v_{em2,cen}$ if em; $v_{abs}$ if otherwise $-$; 
$v_{em,red}$ if $+$} 
}}$$
\par\vfill

\clearpage

\parindent 0pt
\parskip 0pt  
\def\tablerule{\noalign{\hrule\smallskip}}
$$\vbox{\halign{# \hfil & # \hfil & \hfil # \hfil & \hfil # \hfil & 
\hfil # \hfil & \hfil # \hfil & \hfil # \cr
\multispan7{\hfil {\bf{Table 2. Parameters derived from emission
line velocities$^a$}}\hfil} \cr
\noalign{\smallskip} \cr \tablerule \cr \tablerule \cr
Line & How & $T_0$ & \hfil $K_{em}$ & ${\gamma}_{em}$  & $\sigma$ & 
$N$\hfil \cr
 & measured$^b$ & $\rm{HJD} - 2449750$ & \kmsec & \kmsec & \kmsec & 
\cr \noalign{\smallskip} \tablerule \cr
\noalign{\smallskip}
He~I $\lambda\,$5876~\AA\ & Vem,cen &
$5.519 \pm 0.001$ & $85 \pm 4$ & $-60 \pm 3$ & 46 & 327 \cr
\noalign{\smallskip}
He~I $\lambda\,$5876~\AA\ & Vem,blue &
$5.524 \pm 0.002$ & $77 \pm 6$ & $-366 \pm 4$ & 63 & 228 \cr
H$\alpha$ & dgau &
$5.50 \pm 0.1$ & $101 \pm 27$ &  $-34 \pm 15$ & 73 & 329 \cr
\noalign{\smallskip}
H$\alpha$ & Vem,cen$^c$ & 
$5.392 \pm 0.001$ & $72 \pm 3$ & $-94 \pm 2$ & 42 & 329 \cr
\noalign{\smallskip}
H$\alpha$ & Vem,blue &
$5.507 \pm 0.006$ & $83 \pm 15$ & $-675 \pm 13$ & 120 & 88 \cr
\noalign{\smallskip}
He~I $\lambda\,$6678~\AA\ & Vem,cen &
$5.504 \pm 0.008$ & $210 \pm 35$ & $-128 \pm 24$ & 120 & 329 \cr
\noalign{\medskip} \tablerule \cr \noalign{\hrule} \cr
\noalign{\vskip 1pt}
\noalign{$\strut^{\rm a}\ $All fits, by least squares, took the form $V(t)
= \gamma_{em} + K_{em} \sin[2\pi(t-T_0)/P_{orb}].$}
\noalign{For all, the orbital period $(P_{orb})$ was held fixed at
0.1533~d (Patterson et al.~1996).}
\noalign{Also listed are $\sigma,$ the standard deviation of the fit to a
sinusoid, and $N,$ the}
\noalign{number of 60-s summed exposures from both nights, used for each
fit.} \noalign{\vskip 2pt}
\noalign{$\strut^{\rm b}\ $See Section 3.2 and Figure 7 for descriptions
of the line measurements.}
\noalign{\vskip 2pt}
\noalign{$\strut^{\rm c}\ $Adopted for all ephemerides used throughout the
paper}
}}$$
\par\vfill

\end{document}